\documentclass[pre,twocolumn,floatfix]{revtex4}
\usepackage{amssymb}
\usepackage{graphicx}
\usepackage{bm}

\begin{document}

\title
{
Lagrangian statistics and coherent structures in two-dimensional turbulence
}

\author{Michael K. Rivera}
\affiliation{Materials Science \& Technology Division, Los Alamos National
Laboratory, Los Alamos, NM 87545}
\affiliation{Center for Nonlinear Studies, Los Alamos National Laboratory, Los
Alamos, NM 87545}

\author{W. Brent Daniel}
\affiliation{Materials Science \& Technology Division, Los Alamos National
Laboratory, Los Alamos, NM 87545}
\affiliation{Center for Nonlinear Studies, Los Alamos National Laboratory, Los
Alamos, NM 87545}

\author{Robert E. Ecke}
\affiliation{Materials Science \& Technology Division, Los Alamos National
Laboratory, Los Alamos, NM 87545}
\affiliation{Center for Nonlinear Studies, Los Alamos National Laboratory, Los
Alamos, NM 87545}

\begin{abstract}
Measurements of Lagrangian single-point and multiple-point statistics in a quasi-two-dimensional stratifed layer system are reported. The system consists of a layer of salt water over an immiscible layer of Fluorinert and is forced electromagnetically so that mean-squared vorticity is injected at a well-defined spatial scale. Simultaneous cascades develop in which enstrophy flows predominately to small scales whereas energy cascades, on average, to larger scales. Lagrangian velocity correlations and one- and two-point displacements are measured for random initial conditions and for initial positions within topological centers and saddles. The behavior of these quantities can be understood in terms of the trapping characteristics of long-lived centers, the slow motion near strong saddles, and the rapid fluctuations outside of either centers or saddles.
\end{abstract}

\maketitle

\section{Introduction \label{sec: Introduction}}

There is a growing amount of experimental and numerical evidence indicating that two-dimensional (2D) turbulence at moderate to low Reynolds number is dominated by the existence of coherent structures\cite{Bracco-PhF-2000,Farge-PhF-1999}. This observation is more or less independent of whether the turbulence maintains an energetically steady state with a fully developed inverse energy cascade or merely coarsens with time as in decaying 2D turbulence. Coherent structure dominance and the unique aspects of the 2D dual cascade picture, {\em i.e.} an inverse cascade moving energy from small to large length scales and a direct cascade moving mean-squared vorticity, or enstrophy, from large to small length scales, clearly distinguishes 2D turbulence from its three-dimensional (3D) counterpart. An emphasis in the study of 2D turbulence is to understand what impact these structures have on the statistics of turbulence. Recent numerical and experimental results demonstrate that statistical quantities display different behavior in regions around locally hyperbolic points in the velocity field (saddles) and within vortices (centers)\cite{Provenzale-CSF-1995,Daniel-PRL-2002}.  
 
The existence of coherent structures is expected to play an important role in the turbulent dispersion (mixing) of passive scalars\cite{Elhmaidi-JFM-1993}. In particular, centers, if long lived, should act as trapping regions that prevent scalar advection over their encircling manifolds. On the other hand, saddles should quite effectively enhance the dispersion by the stretching of fluid elements into thin filaments along the saddle’s unstable manifold. It is this picture that underlies the description of trajectories of fluid elements in 2D turbulence in terms of successive trapping and flight events, i.e., Levy flights\cite{Shlesinger-PRL-1987}. Although there is little evidence supporting this idea, the proposal highlights the importance of accounting for coherent structures in any mixing model.  
 
How coherent structures affect the mixing properties of 2D turbulent flows can have profound impact on real world problems. An example of geophysical significance is the Antarctic circumpolar vortex which, during the winter months, prevents ozone-depleted air over the South Pole from mixing with ozone-rich air at mid latitudes\cite{Dahlberg-JGRA-1994}. Also, the possibility that 3D turbulence may not be as dominated by coherent structures as is 2D turbulence does not mean that coherent structures do not play an important role in 3D mixing\cite{Lesieur-Turbulence-1987}. Understanding how coherent structures influence mixing in 2D may aid in the understanding of 3D mixing.  
 
The mixing problem is generally approached by applying statistical treatments to fluid-element (or particle) trajectories, i.e., to the Lagrangian dynamics of the fluid. Classical descriptions of Lagrangian dynamics involve the statistical properties of both single trajectories ($N = 1$) following the work of G.I. Taylor\cite{Taylor-PLMS-1921} and relative displacement of two trajectories ($N = 2$) following the work of Richardson\cite{Richardson-PRSL-1926}. In this manuscript, the effects of coherent structures on mixing in 2D turbulence will be quantified by measuring their effects on these classical quantities. Recently there has been important theoretical and numerical work on the evolution of fluid “patches” (groups of $N$ fluid elements with $N>2$)\cite{Falkovich-RMP-2001}. This approach is interesting and could include the effects of coherent structures on the evolution of fluid patches but is beyond the scope of the present work.  
 
Following the discussion in \cite{Falkovich-RMP-2001}, the classical descriptions of particle motion in 2D fluids start from the observation that the Lagrangian trajectory, ${\bf r} \equiv {\bf r}(t)$, of a fluid element or non-inertial particle embedded in the fluid is governed by the stochastic equation:
\begin{equation}
d{\bf r} = {\bf u}({\bf r},t)dt + 2(2 \kappa)^{\frac{1}{2}}\beta(t).
\label{eq: StochasticEquation}
\end{equation}
In Eq.~\ref{eq: StochasticEquation}, ${\bf u}({\bf r},t)$ is the incompressible fluid velocity at position ${\bf r}$ at time $t$, $\kappa$ is the molecular diffusivity and $\beta$ represents standard Brownian motion.  Since our system is in a regime dominated by turbulent advection and not by thermal fluctuations, $\kappa$ will be set to zero.

For single-particle trajectories, the important statistical property is the distance a particle has traveled, on average, over a given time interval $\delta t \equiv t-t_0$.  Defining $\delta {\bf r} \equiv {\bf r}(t) - {\bf r}(t_0)$ one can show that
\begin{equation}
\frac{d}{dt} (\delta {\bf r})^2 = 2 {\bf u}({\bf r}(t),t) \cdot \int_{t_0}^{t}
{\bf u}({\bf r}(s),s) ds,
\end{equation}
which, upon performing an ensemble average and assuming ${\bf u}$ is statistically stationary, becomes an equation for the second moment of the single particle displacement
\begin{equation}
\frac{d}{dt} \langle (\delta {\bf r})^2 \rangle = 2 \int_{t_0}^{t} \langle {\bf u}({\bf r}(t_0),t_0) \cdot {\bf u}({\bf r}(s),s) \rangle ds.
\label{eq: SinglePointIntegral}
\end{equation}
The behavior of $\langle (\delta {\bf r})^2 \rangle$ depends on how $\delta t$ compares with the Lagrangian correlation time $\tau_L \equiv \int_{t_0}^{\infty} \langle {\bf u}({\bf r}(t_0),t_0) \cdot {\bf u}({\bf r}(s),s) \rangle ds/\langle{\bf u}^2\rangle$. If $\delta t \ll \tau_L$ then ${\bf u}({\bf R}(t),t) \approx {\bf u}({\bf r}(t_0),t_0)$ so that the term within the integral of Eq.~\ref{eq: SinglePointIntegral} simply becomes $\langle {\bf u}^2 \rangle$.  In this case $\langle (\delta {\bf r})^2 \rangle \approx \langle{\bf u}^2\rangle \delta t^2$, {\em i.e.} ballistic motion.  If, on the other hand, $\delta t \gg \tau_L$ then the integral in Eq.~\ref{eq: SinglePointIntegral} reduces to $\tau_L \langle {\bf u}^2 \rangle$ yielding Brownian walk behavior for the single particle displacement: $\langle (\delta {\bf r})^2 \rangle \approx 2 \langle{\bf u}^2\rangle \tau_L \delta t$.  These limits follow simply from the properties of the velocity autocorrelation.  For intermediate times the behavior of $\langle (\delta {\bf r})^2 \rangle$ depends on the nature of the inertial turbulent fluctuations and not much is known about its scaling in 2D turbulence\cite{Elhmaidi-JFM-1993}.

For two-particle trajectories, the quantity of interest is the statistics of the relative displacement of particles in a turbulent flow.  Defining the relative displacement of two trajectories ${\bf R}_{12} = {\bf r}_2(t) - {\bf r}_1(t)$, Eq.~\ref{eq: StochasticEquation} implies that
\begin{equation}
\frac{d}{dt}{\bf R}_{12} = {\bf u}({\bf r}_2,t) - {\bf u}({\bf r}_1,t) \equiv
\Delta {\bf u}. 
\end{equation}
As in the case of the single trajectories, an equation for ${\bf R}^2$ can be obtained (dropping the subscripts for simplicity):
\begin{equation}
\frac{d {\bf R}^2}{dt} = 2 {\bf R}\cdot\Delta{\bf u}.
\end{equation} 
If ${\bf R}$ is small, $\Delta {\bf u}$ should be linearly proportional to ${\bf R}$, {\em i.e.}, the first term in a Taylor series.  This linear behavior yields exponential growth in time for ${\bf R}^2$.  On the other hand, if ${\bf R}$ is within the inertial energy range one expects that $\Delta{\bf u} \cdot \hat{R} \approx |{\bf R}^{1/3}|$\cite{Frisch-Turbulence-1995}.  This yields $ |{\bf R}|^{2/3}-R_0^{2/3} \approx \delta t $ where $R_0$ is the trajectory spacing at time $t_0$.  Under the assumption that $R_0$ is small compared to the scales of interest we get $|{\bf R}|^2 \approx (t-t_0)^3$, the famous $t^3$ law derived by Richardson\cite{Falkovich-RMP-2001}.  A Richardson range was reported in both 2D\cite{Jullien-PRL-1999} and 3D experimental systems\cite{Ott-JFM-2000}.
  
Our work on 2D Lagrangian statistics and coherent structures begins with a description and characterization of the experimental apparatus and measurement techniques in Sec.~\ref{sec: Experimental}.  Section \ref{sec: Lagrangian} presents the basic single trajectory and two trajectory Lagrangian statistics.  In Sec.~\ref{sec: CoherentStructures} the quantities presented in Sec.~\ref{sec: Lagrangian} are conditioned on the presence of coherent structures to elucidate the role that such structures play in the mixing process. Finally, in Sec.~\ref{sec: Conclusions}, the implications of and future opportunities for these Lagrangian measurements of turbulence in 2D systems are discussed.

\section{Experimental Methods \label{sec: Experimental}}

Section~\ref{subsec: System} describes the stratified electromagnetically
forced layer that was the quasi 2D fluid used in this set of experiments, as
well as the technique used for obtaining velocity and trajectory information,
namely particle tracking.  In Sec.~\ref{subsec: BPHK} the technique for
identifying coherent structures within the flow is established.  Finally,
Sec.~\ref{subsec: Characterization} presents basic statistical properties of
the turbulence produced in the layer.

\subsection{System and measurement \label{subsec: System}}

The two-point prediction of $\langle |{\bf R}|^2 \rangle \sim t^3$ obtained in Sec.~\ref{sec: Introduction} requires the presence of an inverse energy cascade where $\langle \delta u_{||}^2 \rangle$ scales as $R^{2/3}$ (as we will see this is not a strict requirement).  For 2D, or more pointedly quasi-2D, turbulence this implies that the turbulence must be continuously forced.  One method for continuously forcing the body of a fluid, originally pioneered by Dolzhansky in 1979 \cite{Bondarenko-FAO-1979}, is to subject a current carrying fluid to external magnetic fields.  This method of forcing has since evolved into the stratified electromagnetic layer\cite{Paret-PRL-1997} which has become one of the more common systems for the study of 2D turbulence.

\begin{figure}[b]
\includegraphics[width=3.5in]{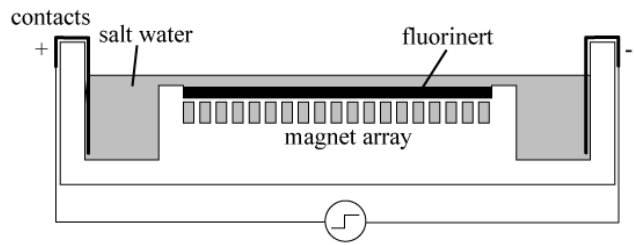}
\caption{\label{fig: Apparatus} Schematic illustration of the stratified layer 2D turbulence experiment. See Sec.~\ref{subsec: System} for detailed description.}
\end{figure}
  
The stratified electromagnetic layer is comprised of a dense salt water layer, typically of order $0.3$ cm deep and $20$ cm $\times$ $20$ cm square, underneath a less dense solution of roughly the same depth.  A current is passed in the plane of the layer and the layers are subject to a spatially varying magnetic field penetrating them vertically.  The resultant Lorentz force drives fluid motion.  The evolution of the stratified layer system is expected to approximate the forced/damped 2D Navier-Stokes equation,
\begin{equation}
\frac{\partial u_i}{\partial t} + u_s\frac{\partial u_i}{\partial x_s} = -
\frac{\partial p}{\partial x_i} + \nu \frac{\partial^2 u_i}{\partial x_i^2} -
\alpha u_i + F_i,
\label{eq: NavierStokes}
\end{equation}
supplemented by the incompressibility condition $\partial_i u_i = 0$.  As usual, ${\bf u}$ is the fluid velocity field, $p$ is the density normalize pressure, ${\bf F}$ is the external electromagnetic forcing, $\nu$ is the fluid kinematic viscosity, and Einstein summation is used throughout.  The linear term with coefficient $\alpha$ represents the effects of frictional drag owing to the container bottom.

In the set of experiments presented here, the system described above is modified slightly by replacing the lower layer of fluid with Fluorinert FC-75 and the upper layer by a dense salt water solution of $20$\% by mass NaCl with a small amount of liquid detergent added to lower surface tension and help with dissolution of tracer particles.  The Fluorinert is used because it has a density $1.8$ times that of water with near the same viscosity, which allows for much stronger stratification than in the case of two salt water solutions. It is also a strong dielectric so the lower layer is completely passive and only the upper salt water layer is electromagnetically forced.  Finally, and perhaps most importantly, the Fluorinert and water are immiscible.  These combined features allow the Fluorinert system to maintain stratification indefinitely, allowing the salt water layer to be driven harder than in previous systems \cite{Paret-PRL-1997}.

\begin{figure}[t]
\includegraphics[width=3.5in]{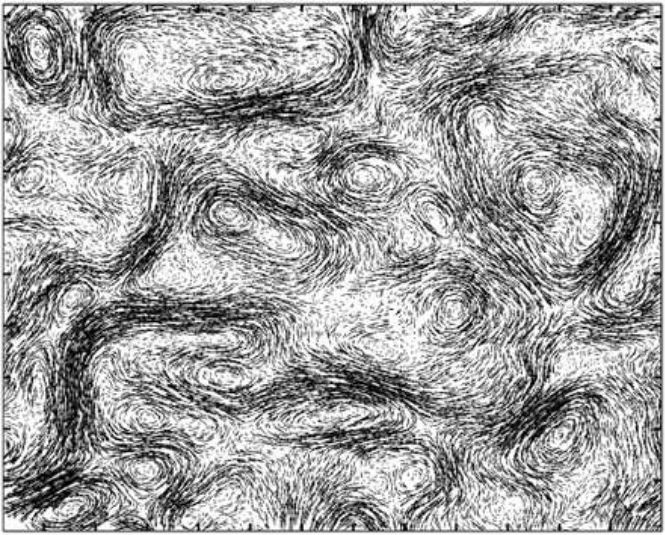}
\caption{\label{fig: RawTracks} Raw particle tracks obtained from the stratified layer over four consecutive frames ($0.07$ s).}
\end{figure}

The experimental apparatus, shown schematically in Fig.~\ref{fig: Apparatus}, consisted of a $0.3$ cm thick layer of salt-solution suspended over a $0.3$ cm thick layer of Fluorinert.  The layers were contained in a $40$ cm $\times$ $20$ cm box with reservoirs at each end across which copper electrodes were placed in the fluid. For the results reported in this paper an alternating square-wave current with frequency $0.5$ Hz and amplitude $0.75$ amps was driven through the salt solution.  A set of $1.27$ cm diameter rare-earth magnets of approximately $0.7$ T residual field strength were arranged with alternating field direction in a $20$ cm $\times$ $20$ cm square array with a period of $2.54$ cm and oriented at $45^\circ$ with respect to the current direction.  The combination of the current and the magnetic field produces a Kolmogorov-like forcing\cite{Bondarenko-FAO-1979,Rivera-PRL-2003} of alternating shear bands with the shear direction along ${\hat y}$ and periodicity $r_{inj} \approx 1.8$ cm in the ${\hat x}$ direction (which implies $k_{inj} \equiv 2 \pi/r_{inj} = 3.5$ rad/cm).  Using the layer depths of $0.3$ cm yields an approximation of $\alpha \approx 0.125$ Hz for the frictional coupling assuming a simple linear shear in the Fluorinert.  Also, the salt solution upper layer has a viscosity around $1.15$ that of water.

To obtain velocity information as well as trajectory information used in generating Lagrangian statistics, the upper salt-solution layer was seeded with polycrystalline powder with mean diameter $75$ $\mu m$ and density $0.98$ gm/cc. Images of the particle fields, illuminated using several Xenon short-arc flash lamps, were obtained with a $1280$ $\times$ $1024$ pixel CCD camera at a frame rate of $60$ Hz.  The velocity field was obtained from image pairs using particle tracking velocimetry derived from two earlier methods\cite{Ishikawa-MST-2000,Ohmi-EiF-2000}.  For a typical pair of images, of order $2.5 \times 10^4$ particle tracks were obtained which were interpolated to a $126 \times 100$ velocity field array.  A typical raw particle track field is shown in Fig.~\ref{fig: RawTracks}.

\begin{figure}[b]
\includegraphics[width=3.5in]{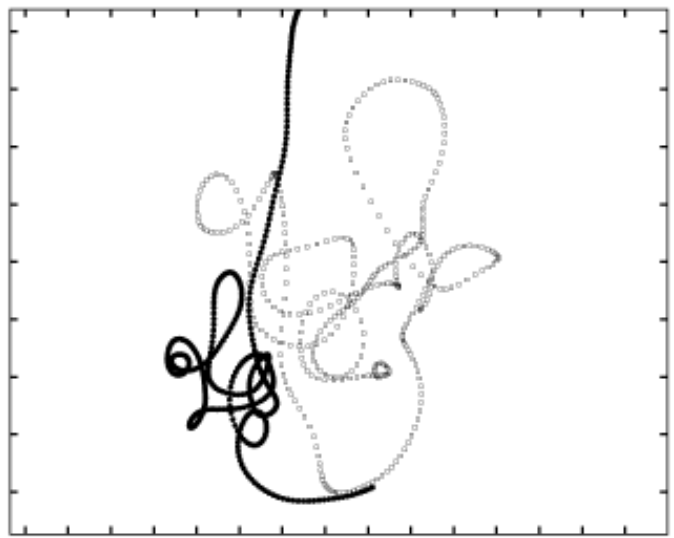}
\caption{\label{fig: Trajectories} Typical time-spliced real particle trajectory ($\Box$) and trajectory computed from dynamic velocity field data ($\bullet$).}
\end{figure}

From this raw data one can obtain the required particle trajectories in two ways: splice together the raw particle tracks or use the interpolated Eulerian fields and generate particle trajectories by solving the advection equation. Both techniques will be used here.  Most of the results will rely, however, on the generated tracks because it is difficult to obtain enough statistics with particle track splicing (an error rate of only $1$ percent splicing from one frame to the next quickly grows as more time steps are taken).  Generated particle tracks also allow for arbitrary initial conditions, a great advantage when exploring the behavior of coherent structures.  The generated particle tracks solved the equation
\begin{equation}
\frac{d {\bf r}}{d t} = {\bf u}({\bf r}(t),t),
\end{equation}
where ${\bf r}(t)$ is the position of the tracer particle at time $t$, using bicubic interpolation to approximate ${\bf u}$ at the particle position and fourth-order Runga Kutta to perform the time integration.  Typical time-spliced particle trajectories and generated trajectories are shown in Fig.~\ref{fig: Trajectories}.

\subsection{Coherent structure identification: the BPHK technique \label{subsec: BPHK}}

As stated in Section \ref{sec: Introduction}, we want to understand how the statistics of the Lagrangian dynamics depend on the existence of coherent structures, such as vortices and saddles.  We use here an Eulerian coherent structure census techniques to \emph{approximate} the coherent structure regions in the flow using the interpolated velocity fields obtained through particle tracking.  The BPHK criterion \cite{Basdevant-PhD-1994,Hua-PhD-1998} for the existence of a coherent structure is a constrained form of the Okubo-Weiss criterion \cite{Okubo-DSR-1970,Weiss-PhD-1991}.  For this reason, following the presentation given in \cite{Basdevant-PhD-1994}, we start with a brief description of Okubo-Weiss. Assume that we are dealing with an inviscid 2D fluid.  Given this assumption, the time evolution of the vorticity field is given by
\begin{equation}
\frac{d \omega}{dt} = \frac{\partial \omega}{\partial t} + u_i\frac{\partial
  \omega}{\partial x_i} = 0.
\label{eq: LagrangianEuler}
\end{equation}
The equation simply states that the vorticity following a fluid parcel is
conserved in Euler fluids.  Consider, for the same fluid, the evolution of the
vorticity gradient, $g_i \equiv \partial_i \omega$, also following the path of
the fluid.  The equation of motion for ${\bf g}$ is:
\begin{equation}
\frac{d g_i}{dt} + A_{ij}g_j=0,
\label{eq: LagrangianGradient}
\end{equation}
where the the velocity gradient tensor is given by $A_{ij}\equiv \partial_i
u_j$.

Assume that the matrix ${\bf A}$ evolves slowly compared to the vorticity gradient ${\bf g}$.  Then one may consider ${\bf A}$ as essentially constant, and the vorticity gradient's evolution is determined locally by the velocity gradient.  Since ${\bf A}$ is traceless by incompressibility, the local topology is determined by $|{\bf A}|$.  In the case where $|{\bf A}|>0$, the eigenvalues of ${\bf A}$ are imaginary and the vorticity gradient is subject to a rotation.  Otherwise, the eigenvalues are real and the vorticity gradient is subject to a strong strain, which enhances the vorticity gradient.  Thus within the Okubo-Weiss approximation there are two types of topological structure within the flow, centers for which $|{\bf A}| \gg 0$ and saddles for which $|{\bf A}| \ll 0$.
  
The critical assumption in Okubo-Weiss, as pointed out by Basdevant and Philipovitch \cite{Basdevant-PhD-1994}, is that the velocity gradient matrix evolves slowly compared with the vorticity gradient.  This assumption can be made more quantitative by taking the material derivative of Eq. \ref{eq:   LagrangianGradient}.  This yields
\begin{equation}
\frac{d^2 g_i}{dt^2} + g_j\frac{dA_{ij}}{dt} + A_{ij}\frac{d g_j}{dt}=0.
\label{eq: SecondLagrangianGradient} 
\end{equation}
For the Okubo-Weiss assumption to be true, the second term must be much smaller than the third.  That is
\begin{equation}
\left( \frac{|g_j\frac{dA_{ij}}{dt}|}{|A_{ij}\frac{d g_j}{dt}|} \right) \ll 1.
\label{eq: BPHK-raw}
\end{equation}
Without tracing the algebra (which is clearly enumerated in either \cite{Basdevant-PhD-1994} or \cite{Hua-PhD-1998}), the above inequality can be re-written in terms of the eigenvalues $\lambda_{\pm}$ of the Hessian of pressure, $H_{ij}\equiv\partial^2_{ij} p$.  Using these, Eq. \ref{eq: BPHK-raw} becomes
\begin{equation}
R\equiv\left( \frac{(\lambda_+ - \lambda_-)^2}{(\lambda_+ + \lambda_-)^2} \right) \ll 1.
\end{equation}

\begin{figure}
\includegraphics[width=3.5in]{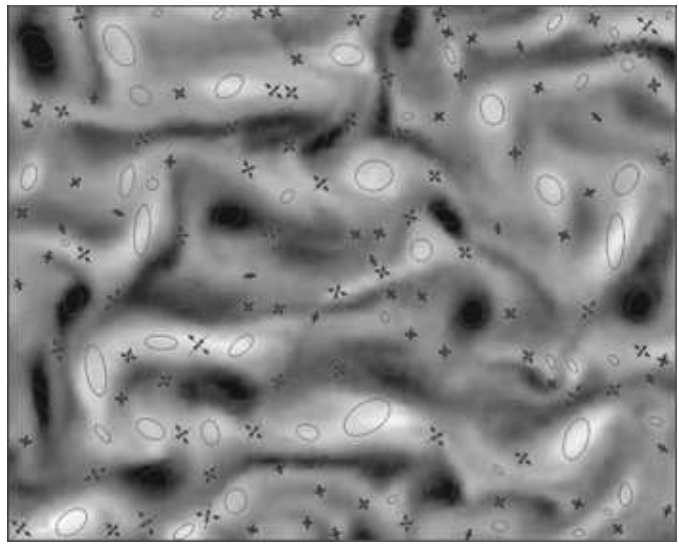}
\caption{\label{fig: Vorticity} BPHK coherent structure regions superimposed on the corresponding vorticity field.  Ovals enclose centers (or vortices) and crosses denote the saddle positions.}
\end{figure}

The implementation of BPHK is now fairly straight-forward.  From the measured velocity fields evaluate the pressure field by inverting $\nabla^2 p = 2|{\bf A}|$. Obtain the pressure Hessian and evaluate the field $R$.  Find points for which $R<1$ corresponding to positions where the Okubo-Weiss assumption is valid.  Now evaluate $|{\bf A}|$ at these positions and separate them into regions for which $|{\bf A}| < 0$ (saddles) and $|{\bf A}| > 0$ (centers). These points comprise the coherent structures.  The coherent structure identified using the BPHK criterion superimposed on the corresponding vorticity field is displayed in Fig.~\ref{fig: Vorticity}.

\subsection{System characterization \label{subsec: Characterization}}

\begin{figure}[t]
\includegraphics[width=3.25in]{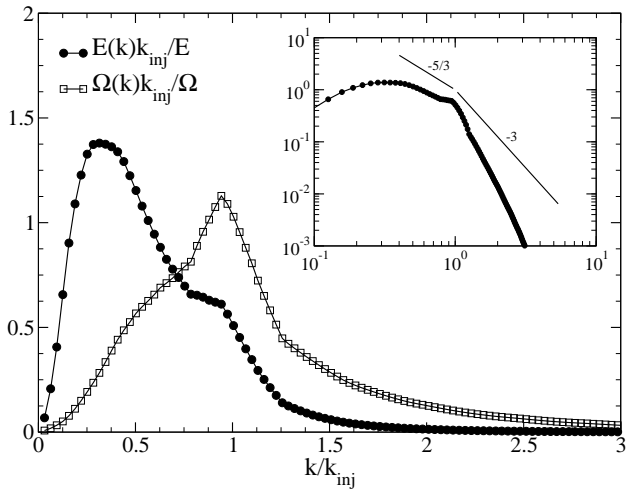}
\caption{\label{fig: Spectra} The energy and enstrophy spectrum obtained from the electromagnetic layer for the driving conditions described in Sec.~\ref{sec: Experimental}.  Inset are the same spectra on a log-log plot.}
\end{figure}

Under the driving conditions and for the magnetic configuration described in Sec.~\ref{subsec: System} the turbulence in the stratified layer system described above can be characterized by several of its Eulerian ensemble-averaged statistics.  The average energy per unit mass in the system $E = u_{rms}^2/2$ was $8.4$ cm$^2/$s$^2$ ($u_{rms}=4.1$ cm/s) and the average enstrophy $\Omega = \omega_{rms}^2/2$ where the vorticity $\omega \equiv {\bf \nabla} \times {\bf u}$ was $51$ s$^{-2}$ ($w_{rms}=10.1$ s$^{-1}$).  The spectra of energy $E(k)$ and enstrophy $\Omega(k)$ calculated using a Welch spatial window are shown in Fig.~\ref{fig: Spectra}.  There is a build up of energy at large scales as expected for a system with an inverse cascade of energy.  However, the spectra never achieve the $k^{-5/3}$ for $E(k)$ and $k^{1/3}$ for $\Omega(k)$ consistent with the existence of an energy inertial range.  Also note that the spectra are significantly steeper in the enstrophy range than the predicted $k^{-3}$, possibly due to the three dimensionality of the system at small scales.  Using the spectra we can obtain the outer scale approximation of $L_{out} \approx 6$ cm and injection scale velocity, $u_{inj}=1.8$ cm/s and vorticity $\omega_{inj}=5.8$ s$^{-1}$.

The energy dissipated by viscosity and the frictional damping are given by $\epsilon_{\nu} = \nu \omega_{rms}^2$ and $\epsilon_{\alpha} = \alpha u_{rms}^2$ respectively.  Using the coefficients $\alpha$ and $\nu$ given in Sec.~\ref{subsec: System} yields $\epsilon_{\nu}=1.2$ cm$^2$/s$^3$ and $\epsilon_{\alpha} = 2.1$ cm$^2$/s$^3$.  Since the system is in an energetically steady state this implies that the energy injection rate $\epsilon_{inj} = 3.3$ cm$^2/$s$^3$.  That more energy is lost to friction is consistent with an inverse cascade that moves energy predominantly away from small scales where viscous dissipation is efficient.  One can also use the estimate $\epsilon_{inj} \approx u_{inj}^{3}/r_{inj}$ which yields a value of $2.9$ cm$^2/$s$^3$, consistent with the above value.  Similar results can be obtained for the enstrophy injection and dissipation rates.  For these we obtain $\beta_{inj}=39.2$ s$^{-3}$, $\beta_{\alpha}=12.3$ s$^{-3}$ and $\beta_{\nu}=26.9$ s$^{-3}$.

\begin{figure}[b]
\includegraphics[width=3.25in]{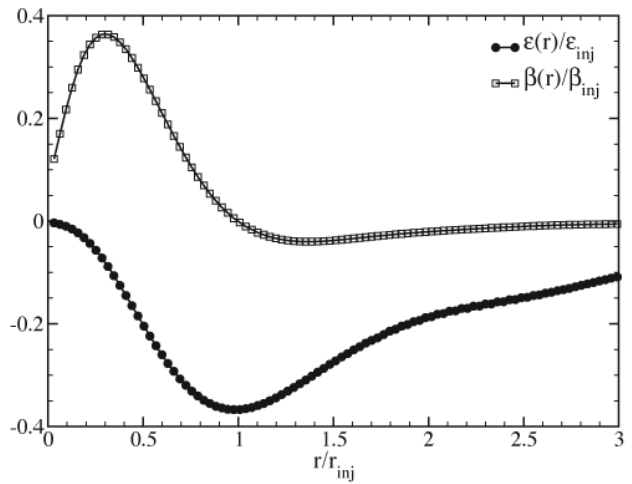}
\caption{\label{fig: EnergyEnstrophyTransfer} The scale-to-scale transfer of energy and enstrophy obtained from the electromagnetic layer for the driving conditions described in Sec.~\ref{sec: Experimental}.}
\end{figure}

The bulk dissipation rates can be further characterized by measuring the transfer of energy $\epsilon(r)$ and enstrophy $\beta(r)$ through a length scale $r$ using the filter approach reported in \cite{Rivera-PRL-2003}.  These functions are displayed in Fig.~\ref{fig: EnergyEnstrophyTransfer}.  The flux directions are correct, predominantly downscale for enstrophy and upscale for energy. It should be stressed that the choice of filter used in the filter approach, namely a Gaussian, has a tendency to smooth features; different filters would produce somewhat flatter regions ({\em i.e.} more inertial) in the energy and enstrophy transfer and sharper behavior around the energy injection scale.  Use of such filters, however, is prohibited by the existence of measurement boundaries.

Finally, several length and time scales can be extracted from the reported coefficients.  The Kolmogorov scale $\eta=(\nu^3/\beta_{inj})^{1/6}\approx 0.07$ cm and the Taylor scale $\lambda=(E/\Omega)^{1/2} \approx 0.41$ cm, from which we can obtain the Taylor scale Reynolds number $Re_{\lambda}=\lambda u_{rms}/\nu = 112$. Similarly the injection scale and outer scale Reynolds numbers are $Re=r_{inj}u_{inj}/\nu = 303$ and $Re_o = L_o u_{rms}/\nu = 2150$.  We should note that $\eta$ is small, much less than the salt layer depth of $0.3$ cm.  We therefore expect 3D effects to become important before the Kolmogorov scale is reached.  Characteristic times that one can define for the turbulent state are the large-eddy turnover time $\tau_o = L_o/u_{rms} = 1.5$ s, the injection scale eddy turnover time $\tau_{inj} = r_{inj}/u_{inj} = 1$ s and a characteristic shear time $\tau_\Omega = \Omega^{-1/2} = 0.14$ s.

\section{Lagrangian statistics \label{sec: Lagrangian}}

Before presenting results involving coherent structures, Lagrangian single- and two-particle statistical quantities for the full data fields are considered.  In this section both real and generated particle trajectories are used in the analysis.

\subsection{Single trajectory statistics: Self-diffusion \label{subsec: SelfDiffusion}}

\begin{figure}[t]
\includegraphics[width=3.25in]{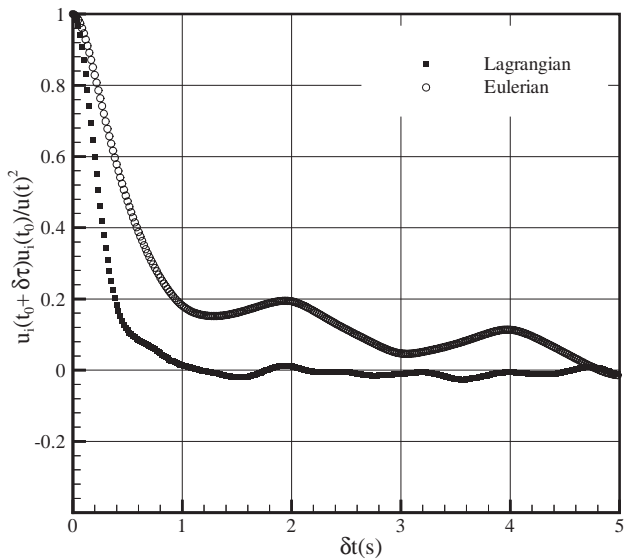}
\caption{\label{fig: TimeCorrelation} The Lagrangian and Eulerian velocity correlation function obtained using real particle tracks from the stratified layer.  Using these functions the Lagrangian correlation time (see Sec.~\ref{sec: Introduction} for definition ) is found to be $\tau_L=0.26$ s. Similarly the Eulerian correlation time is $\tau_E = 0.95$ s.}
\end{figure}

Single trajectory statistics depend on how the time increment $(t- t_0)$ compares with the Lagrangian correlation time $\tau_L \equiv \int_{t_0}^{\infty}\langle {\bf u}({\bf r}(t_0),t_0) \cdot {\bf u}({\bf r}(s),s) \rangle ds / \langle |{\bf u}|^2 \rangle$. Figure \ref{fig: TimeCorrelation} shows the velocity autocorrelation function for both Lagrangian trajectories and Eulerian points. The latter is displayed to contrast the 2D turbulence of the stratified layer with 3D turbulence results where $\tau_E < \tau_L$ ($\tau_E$ has the same definition as $\tau_L$ except that the ``trajectory'' is a constant point in space). Performing the integration and normalization yields $\tau_L = 0.26$ s and $\tau_E = 0.95$ s.  The ratio $\tau_E / \tau_L = 3.7$ is consistent with earlier numerical simulations of 2D turbulence which found the ratio to have values between $3$ and $4$ \cite{Elhmaidi-JFM-1993}. The oscillations in the Eulerian autocorrelation function, corresponding to the $0.5$ Hz square wave in the electromagnetic driving current, are not apparent in the Lagrangian autocorrelation function because the particle trajectories randomly sample spatial points, averaging out the oscillations.

\begin{figure}[b]
\includegraphics[width=3.5in]{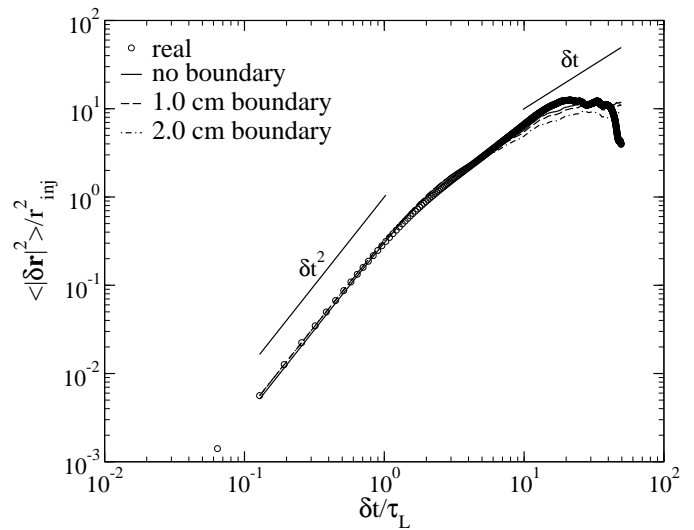}
\caption{\label{fig: SelfDiffusion} Single-particle Lagrangian displacement $\langle |\delta {\bf r}|^2 \rangle$ for real and generated trajectories.  With the generated trajectories several different boundary sizes were utilized to illustrate the effect of finite data extent on the results.}
\end{figure}

The second moment of single particle displacements $\langle |\delta {\bf r}|^2 \rangle$ is displayed in Fig. \ref{fig: SelfDiffusion} as a function of time displacement $(t - t_0)$ for both real and generated trajectories. The agreement between the two is fair, the largest difference being at large scales where imposing artificial boundaries on the generated particles limits the maximum measurable displacement and reduces the average. There is a slight disagreement at approximately $2 \tau_L$ where the generated data tends to overshoot the real data. This may be statistical bias in the particle tracking algorithm (splicing tracks over smooth regions is easier than over heavily strained regions) or could be an indication of inertial behavior of the real particles.

The trend in $\langle |\delta {\bf r}|^2 \rangle$ for $(t- t_0)<L$ is power-law scaling with an exponent $\sim 2$ as expected in the ballistic regime of motion. The amplitude is $14$ cm$^2/$s$^2$, fairly close 
to the expected value of $u^2_{rms} = 17$ cm$^2/$s$^2$.  For $(t - t_0)>L$ the exact exponent is sensitive to the boundary, but a power law fit to the data without imposed boundaries yields an exponent of about $1.3$ in agreement with earlier numerical results\cite{Elhmaidi-JFM-1993}. Finally, no Brownian diffusive regime is reached in the current data set. To see this regime, data from a significantly larger spatial area would be required. 

\subsection{Two trajectory statistics: The Richardson regime \label{subsec: Richardson}}

\begin{figure}
\includegraphics[width=3.5in]{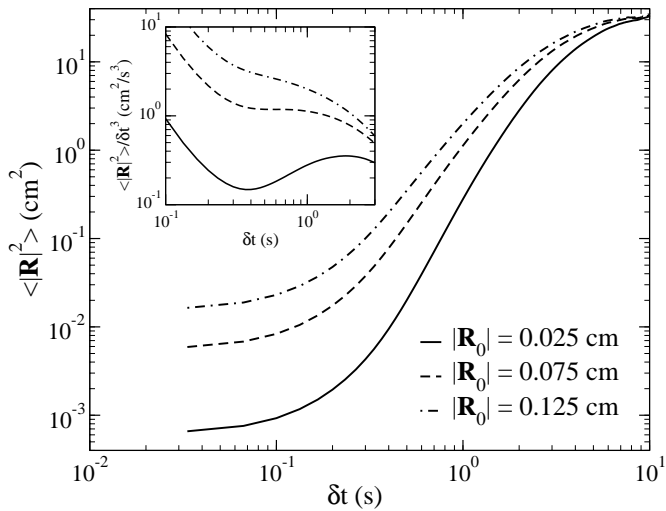}
\caption{\label{fig: Richardson} The mean square relative displacement $\langle |{\bf R}|^2 \rangle$ of two-particle trajectories in the flow for several different initial separations $R_0$.  All of the plots were created using generated particle trajectories.  Inset are the same plots normalized by the Richardson predicted behavior of $\delta t^3$.}
\end{figure}

As discussed in Sec. \ref{sec: Introduction}, the expected behavior of the two trajectory Richardson 
statistics in an inverse cascade inertial range is $\langle |{\bf R}|^2 \rangle \sim \delta t^3$. In Fig. \ref{fig: Richardson} the two-trajectory statistics for $\langle |{\bf R}|^2 \rangle$ are displayed for several initial displacements $|{\bf R_0}|$. A very similar set of curves were obtained from numerical simulations of 2D turbulence \cite{Elhmaidi-JFM-1993}. Inset are the same plots compensated by $(t - t_0)^3$ . It should also be noted that the expected exponential small time behavior (not shown) is not a clean exponential for the generated data. Moreover, the dependence of $\langle |{\bf R}|^2 \rangle$ is markedly different for real and generated data at small time, though long time behavior is identical (real data tends to show more of an $t^2$ scaling at small time possibly due to the real particles inertia). Due to statistical constraints, the generated data will be used in subsequent analysis.

For an initial displacement of $0.075$ cm, there seems to be a nice $\delta t^3$ range from $0.3$ s to $1.0$ s. That the scaling depends on the initial separation and that the spatial scale where scaling is apparently observed is below the injection scale are two reasons to be cautious about interpreting this scaling as indicative of the predicted Richardson result\cite{Jullien-PRL-1999}. Further, one expects crossover effects \cite{Chertkov-communication} to play an important role in the precise determination of scaling for moderate Reynolds number flows. Details  of Richardson scaling and comparisons with results from other experiments\cite{Jullien-PRL-1999} will be presented elsewhere.  Here the effects of conditioning the separation on whether the initial positions of the particles is within saddles or centers is considered. Unless otherwise noted the initial distance of $|{\bf R_0}|=0.75$ cm will be used for the rest of the analysis.

\section{Effects of coherent structures \label{sec: CoherentStructures}}

In this section the statistical quantities measured in Sec. \ref{sec: Lagrangian} for ensembles of particles starting at any given point in the flow are conditioned on whether the initial particle position ${\bf r}(t_0)$ is within a saddle or center as identified using the BPHK technique described in Sec. \ref{sec: Experimental}. All results presented in this section were obtained from generated data rather than real particle tracks so that statistical convergence could be achieved. In general one would like to separate the statistics on the basis of how long that particle was within the coherent structure and how long that coherent structure existed. This separation would be particularly important for the case of saddles which do not have a closed group of core particles associated with them. Unfortunately, such conditions further constrain statistics that have already been pushed to their limit, and thus, only the approach based on initial conditions is possible. 

\subsection{Conditional single trajectory statistics \label{subsec: ConditionalSelfDiffusion}}

\begin{figure}
\includegraphics[width=3.5in]{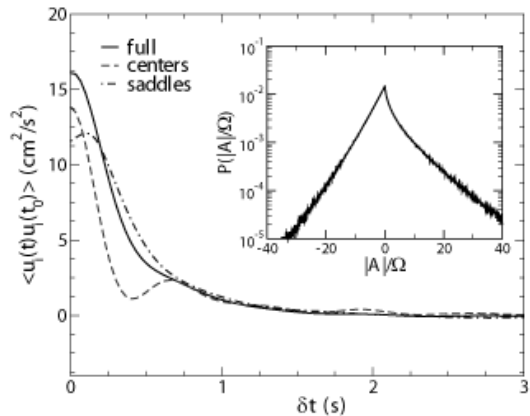}
\caption{\label{fig: CoherentCorrelation} The conditional Lagrangian velocity correlation function obtained using generated particle tracks from the stratified layer.  Dashed and dash-dot lines correspond to conditioning upon the type of structure the trajectory is within at time $t_0$.  The solid line is the unconditioned average.  Inset is the probability distribution for the topological quantity $|{\bf A}|$.}
\end{figure}

As in Sec.~\ref{subsec: SelfDiffusion}, the discussion of single trajectory statistics starts with the single point velocity auto-correlation. The correlations are displayed in Fig.~\ref{fig: CoherentCorrelation} where the average $\langle \rangle$ has been conditioned on the type of coherent structure the particle is within at time $t_0$; the average over the full data field is shown for comparison.
 
For both saddles and centers, the magnitude of the correlation function at $\delta t = 0$ is suppressed compared with the background flow. The suppression is caused by the self-organization of coherent structures which are relatively quiescent compared with the background flow, and thus they have lower RMS fluctuations. One might be surprised that this effect is stronger for saddles than for centers. The difference is due to a bias in the population statistics of saddles and centers\cite{Daniel-PRL-2002,Rivera-PRL-2001}: it is more probable to find a center of significant strength in the flow than a strong saddle. This argument is supported by considering the probability distribution, $P(|\bf{A}|)$ for the topological quantity $|{\bf A}|$ defined in Sec. \ref{subsec: BPHK} and shown in the inset of Fig. \ref{fig: CoherentCorrelation}. There is a larger probability for $|{\bf A}| \gg 0$ than $|{\bf A}| \ll 0$. For more discussion about why this must be the case see \cite{Rivera-PRL-2001}.

For both saddles and centers, when $\delta t \approx 1$ s this initial difference has, for the most part, disappeared and the correlation functions collapse to the correlation function of the full statistics. For centers, the correlation function oscillates with a period of about $0.6$ s, fairly close to the injection scale eddy rotation time, $\tau_{inj} = 1.0$ s, which is expected owing to the periodic rotation of the centers. The steep fall off in the amplitude of the center-correlation occurs for a number of reasons: the stripping of particles from the exterior of vortices, the random walk of vortices through the flow, and a finite vortex lifetime. In spite of these effects, weak oscillations in the correlation function continue out as far as $2.4$ s, indicating some long lived structures in the flow. Using the center-correlation function, the eddy turnover time is defined to be $\tau_e = 0.6$ s, which is close to the dimensionally estimated value of the injection-scale eddy turnover time of $1.0$ s.
 
The behavior of the correlation function for saddles is somewhat surprising in that it initially increases. This increase comes about because particles starting within a saddle are not confined to the saddle, as they are in the cases of centers, but are leaving the saddle to the background. As already indicated, the background has stronger fluctuations when compared with the relatively calm region within the saddle, thus the increase. The crossing point of the saddle-correlation with the full correlation at 
about $0.25$ s is interpreted to be the average time a particle remains within a saddle structure. After this time, out to about $\tau_e$, the saddle correlation exceeds that of the full statistics. This crossover indicates that the particles, once having left the saddle, continue to remember their interaction with the structure for quite some time, effectively surfing the saddle's unstable manifold. 

\begin{figure}
\includegraphics[width=3.5in]{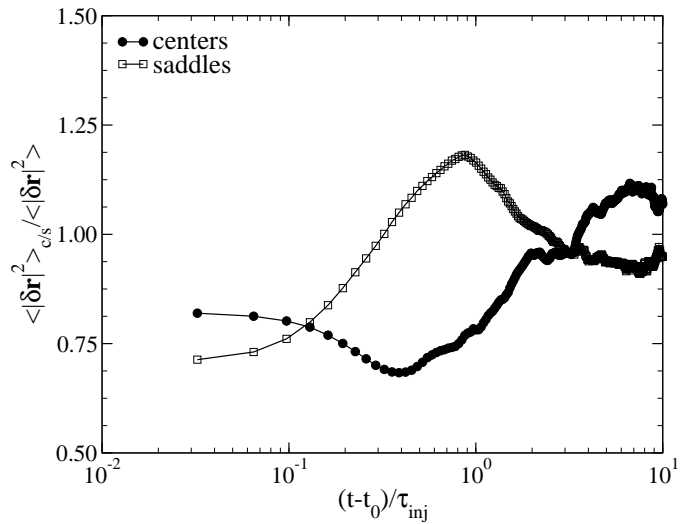}
\caption{\label{fig: CoherentSelfDiffusion} Single-particle Lagrangian displacement $\langle |\delta {\bf r}|^2 \rangle$ for generated trajectories conditioned upon the type of structure the trajectory is within at time $t_0$.  The solid line represents the unconditioned average.}
\end{figure}

Given the observations obtained from the Lagrangian velocity correlation, the conditional single trajectory statistics, displayed along with the unconditioned statistics in Fig.~\ref{fig: CoherentSelfDiffusion}, can be interpreted. One might expect that at small times, corresponding to being within saddle structures, the self diffusion would be enhanced but at very small times self diffusion within saddles is smaller than for any other point in the flow. This slow behavior can be explained by the suppressed RMS fluctuations within saddles resulting from a bias in the population statistics of $|{\bf A}|$. A similar argument explains the suppression of self diffusion within centers at early times. The expected enhancement in self-diffusion within saddles does not occur until times greater than $(t - t_0) =0.25$ s, where the earlier conditional Lagrangian velocity correlation crosses the full Lagrangian correlation function. The crossing time is where most of the trajectories have left the saddle. The conclusion, then, is that saddles enhance diffusion, but non-locally. Only when the particle escapes the confines of the saddle structure and begins to surf the unstable manifold is the enhancement felt. At long times, about $2$ s, the trajectory begins to forget both the saddle and unstable manifold and the conditional statistics approach the full statistics.

The self diffusion for particles beginning within centers behaves more intuitively. The suppression at small times was explained by bias in the statistics of $|{\bf A}|$. At intermediate times, smaller than the eddy rotation time, self-diffusion continues to be suppressed because the trajectories cannot cross the center manifold and continue to circulate about the vortex. Finally, at long times, as the centers dissolve, the conditional statistics approach the full statistics.  Overall, the effect of different initial conditions is not large, accounting for at most a factor of $2$ between centers and saddles. 

\subsection{Conditional two-trajectory statistics \label{subsec: ConditionalRichardson}}

\begin{figure}
\includegraphics[width=3.5in]{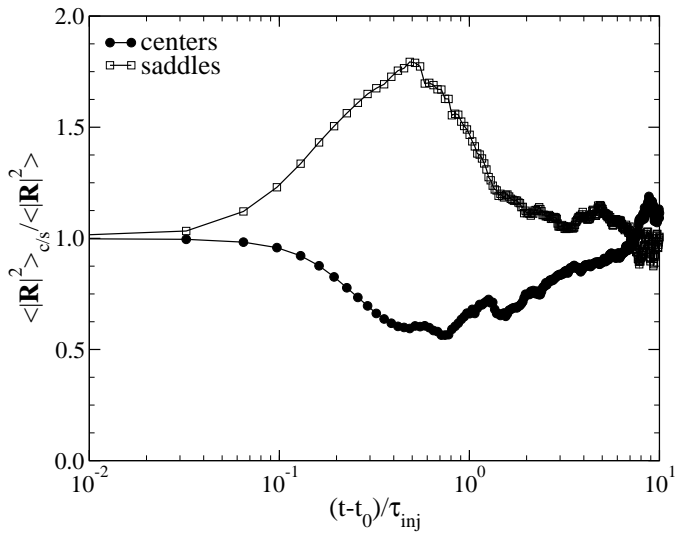}
\caption{\label{fig: CoherentRichardson} Two-particle Lagrangian displacement $\langle |{\bf R}|^2 \rangle$ for generated trajectories conditioned upon the type of structure the trajectory is within at time $t_0$.  The solid line represents the unconditioned average.}
\end{figure}

The results of conditioning the average $\langle |{\bf R}|^2 \rangle$ upon the condition that a trajectory starts within a center or saddle at $t_0$ are displayed in Fig.~\ref{fig: CoherentRichardson} . In these two-trajectory statistics the expected enhanced diffusion due to saddles is realized. From the earliest times to times as long as $2$ s there is an enhancement in the relative diffusion with a peak difference at about $0.25$ s. From Sec. \ref{subsec: ConditionalSelfDiffusion} recall that this earlier time is roughly the time it takes for the particles to leave the saddle structure. Thus, for relative diffusion, in stark contrast to the self-diffusion presented before, saddles behave locally. Once the maximum amount of enhancement occurs, the saddle statistics decay back to the full statistics as the particle pairs forget their origin within the saddle. The decay is slow most likely because some particle pairs initially straddle the stable manifold and thus exit the saddle on opposite sides and surf opposing unstable manifolds. At around $2$ s, however, the statistics collapse back to the full statistics, much as in the case of self-diffusion, because the trajectories begin to forget their origins.

The relative diffusion of particles within centers also behaves, more or less, as expected. The diffusion is suppressed at small times and slowly approaches the full statistics as the vortices are destroyed and the trapped trajectories are released to evolve in the background. One might be surprised that the trapping of vortices does not suppress the diffusion with respect to the full statistics more than the slight amount displayed here. To resolve this issue, consider the types of structures, other than saddles, that exist within the flow. Whatever their form, for example jets, one expects that by definition $|{\bf A}|$ is small, {\em i.e.}, there are small velocity gradients. Thus, in these regions, a pair of particles would evolve together, only separating quickly when they encounter a saddle.
  
\begin{figure}[b]
\includegraphics[width=3.4in]{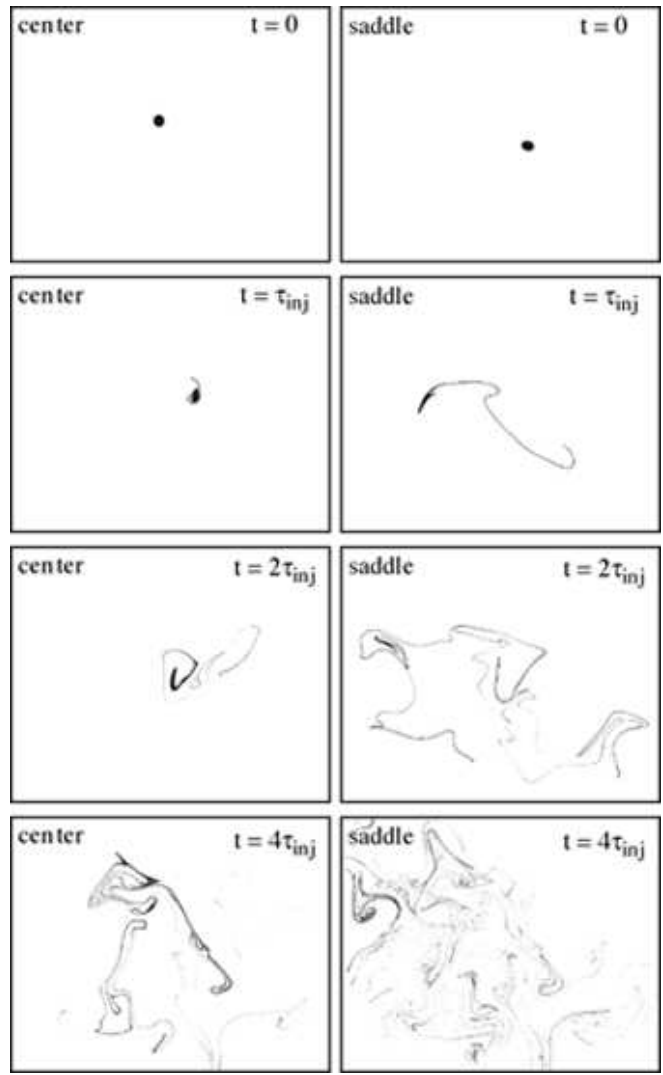}
\caption{\label{fig: ScalarEvolve} The time evolution of an ``ink drop'' of generated particle tracks.  The left column represents the behavior of an ink drop that initially seeds a center region.  The right column is the same for a saddle region.}
\end{figure}

To better visualize the implications of the above results on relative diffusion $10^5$ trajectories were generated from initial positions within a center region and within a saddle. The results of tracking these trajectories over time are displayed in Fig. \ref{fig: ScalarEvolve}. One can see from the figure that the initial spread for trajectories beginning within the saddles is far greater than for the corresponding center region. It should be stressed that only a few of the starting particles within the saddle account for this large spread, the majority continue to exist in the dense grouping of particles in the left of the image. This dense grouping weights statistics, such as those shown in Fig. \ref{fig: ScalarEvolve}, so that they only echo a fraction of the magnitude of the spread.

\section{Conclusions \label{sec: Conclusions}}

The Lagrangian approach of particle diffusion was applied to the quasi-2D turbulence of a thin stratified layer. Simultaneous cascades of enstrophy and energy develop although neither is strictly inertial despite larger Reynolds number and smaller damping than in other stratified layer experiments \cite{Paret-PRL-1997}. Single particle diffusion follows the predictions of Taylor at small times that $\langle \delta {\bf r}^2 \rangle \propto \langle |\bf u|^2 \rangle \delta t^2$ and is consistent with the scaling of $(t -t_0)^{1.3}$ found from numerical simulations for times larger than the Lagrangian velocity correlation time\cite{Elhmaidi-JFM-1993}. The expected long-time behavior of diffusive scaling like $t -t_0$ is not observed owing to the finite size of the system. By separating the statistics based on the initial position within a topological saddle or center, one can understand the influence of these structures on particle diffusion. The overall effect is small with the maximum difference near $\tau_{inj}$ where center dispersion is about half that of particles in the rest of the field.
 
The behavior of two-particle separations is more complex, reflecting the competing effects of crossover between competing scalings at low to moderate Reynolds number and of boundaries. The intermediate time behavior depends on the initial starting separation $R_0$ with decreasing slope for large $R_0$. The existence of a Richardson scaling range with $\langle R^2 \rangle \propto (t -t_0)^3$ is unresolved and will be presented in detail elsewhere. At short times the separation is sensitive to starting conditions, {\it i.e.}, whether the initial mean position was within a saddle or center. As expected, separation increased more rapidly at small times when the particles started in a saddle and less rapidly when they were within a center.

The most interesting observation for the separated statistics is the time scales at which the structures, in particular the saddles, have their greatest effect. For the two-point statistics the greatest effect of structures is felt at time scales for which the particle is within the structure. The single point statistics, on the other hand, are non-local, that is the structures greatest effect on these is at times when the particle has exited the structure. This difference might be explained by the existence of an unaccounted for structure, {\em e.g.} a jet, which lurks on the outside of saddles. When a particle is advected out of the saddle, it is swept up in a jet and advected rapidly away from its starting point, which explains the enhanced self-diffusion at intermediate times. If two particles are advected into a jet they are carried along with little relative displacement, thus the two-point statistics are not enhanced. For this explanation to be compelling, the notion of a jet needs to be quantitatively defined.
  
The potential for the Lagrangian approach to the analysis of turbulence has barely been scratched by these preliminary studies. Multiple-point particle tracking can be implemented easily and would provide a better measure of turbulence transfer processes. Another approach would be to measure interscale transfer flux in the moving frame of a group of particles that form a particular topological structure, {\it e.g.}, saddles or centers. One can compute the stretching fields of the flow to probe the mixing properties of the turbulent system \cite{Voth-PRL-2002}. There are many more analysis ideas that follow along these lines and that take advantage of being able to measure the dynamics of fluid quantities moving with a fluid parcel. The opportunities are numerous and will help elucidate the mechanisms for turbulence and transport in 2D turbulence. Further, the concepts and analysis that can be applied to 2D data will set the stage for similar applications to 3D data when it becomes available.

\section{Acknowledgements}
We acknowledge useful discussions with Misha Chertkov. This work was funded by the U.S. Department of Energy (W-7405-ENG-36).

\end{document}